\documentstyle[11pt,aaspp4,flushrt]{article}

\def\cmm2{{\,\rm cm^{-2}}}
\def\cm2{{\,{\rm cm}^2}}
\def\cmm3{{\,{\rm cm}^{-3}}}
\def\gcmm3{{\,{\rm g\,cm^{-3}}}}

\received{RECEIPT DATE}
\revised{REVISION DATE}
\accepted{ACCEPT DATE}
\cpright{AAS}{1996}

\journalid{VOL}{JOURNAL DATE}
\articleid{START PAGE}{END PAGE}
\paperid{MANUSCRIPT ID}
\cpright{AAS}{1995}
\ccc{CODE}

\lefthead{Sigl et al.}
\righthead{Clustering of Highest Energy Cosmic Rays}

\begin{document}
\vspace*{1cm}
\rightline{FERMILAB-Pub-96/121-A}
\rightline{Submitted to \it Astrophysical Journal Letters}
\vspace{1cm}
\title{IMPLICATIONS OF A POSSIBLE CLUSTERING OF HIGHEST
ENERGY COSMIC RAYS}
\author{G\"unter Sigl$^{1.2}$, David~N.~Schramm$^{1,2}$, \&
Sangjin Lee$^{1,2}$}
\affil{$^1$Department of Astronomy \& Astrophysics\\
Enrico Fermi Institute, The University of Chicago, Chicago, IL 60637-1433\\
$^2$NASA/Fermilab Astrophysics Center\\
Fermi National Accelerator Laboratory, Batavia, IL 60510-0500}
\author{Paolo Coppi$^3$}
\affil{$^3$Department of Astronomy\\
Yale University, New Haven, CT 06520-8101}
\author{Christopher~T.~Hill$^4$}
\affil{$^4$Theoretical Physics, MS 106\\
Fermi National Accelerator Laboratory, Batavia, IL 60510-0500}

\begin{abstract}
Very recently, a possible clustering of a subset of observed ultrahigh
energy cosmic rays above $\simeq40\,$EeV ($4\times10^{19}\,$eV)
in pairs near the supergalactic plane was reported. We show that
a confirmation of this effect would provide
information on origin and nature of these events and, in
case of charged primaries, imply interesting constraints on the
extragalactic magnetic field. The
observed time correlation would most likely rule out an association
of these events with cosmological gamma ray bursts. If no
prominent astrophysical source candidates such as powerful
radiogalaxies can be found, the existence of a mechanism
involving new fundamental physics would be favored.
\end{abstract}

\newpage

\section{Introduction}
The recent detection of ultrahigh energy cosmic rays (UHE CRs)
with energies above $100\,$EeV (Bird et al. 1993,
1994, 1995a; Hayashida et al. 1994; Yoshida et al.
1995) has triggered considerable discussion in the literature on
the nature and origin of these particles (Sigl, Schramm, \&
Bhattacharjee 1994; Elbert \& Sommers 1995; Halzen et al. 1995). 
On the one hand, even the most powerful astrophysical objects
such as radio galaxies and active galactic nuclei (AGN) are
barely able to accelerate charged particles to such energies
(Hillas 1984). On the other hand, above $\simeq70\,$EeV the
range of nucleons is limited by photopion production on the
cosmic microwave background (CMB) to about $30\,$Mpc (Greisen
1966; Zatsepin \& Kuzmin 1966), whereas heavy nuclei are
photodisintegrated on an even shorter distance scale (Puget,
Stecker, \& Bredekamp 1976). In addition, protons above
$100\,$EeV are deflected by only a few degrees over these
distances, if one uses commonly assumed values for the
parameters characterizing the galactic and extragalactic
magnetic fields (Sigl, Schramm, \& Bhattacharjee 1994).

Currently there exist three classes of models for UHE CRs. The
most conventional one assumes first order Fermi acceleration of
protons at astrophysical magnetized shocks (see, e.g. Blandford
\& Eichler 1987). This mechanism is supposed to be associated
with prominent astrophysical objects such as AGN and radio
galaxies. One problem with this scenario is that no obvious
candidate could be found
within a cone around the arrival
direction of the two highest energy events observed whose
opening angle is given by the expected proton deflection angle
(Elbert \& Sommers 1995; Hayashida et al. 1994).

Recently a second class of models has been suggested; namely
that UHE CR could be associated
with cosmological gamma ray bursts (GRBs) (Waxman 1995a,b;
Vietri 1995a; Milgrom \& Usov 1995). This was mainly motivated
by the fact that the
required average rate of energy release in $\gamma$-rays and UHE
CRs above $10\,$EeV turns out to be comparable. Protons could be
accelerated beyond $100\,$EeV within the relativistic shocks
associated with fire ball models of cosmological GRBs (Meszaros
1995). One advantage of such a scenario is that sources are not
necessarily correlated with powerful astrophysical objects. In
fact, the high degree of isotropy observed in the GRB
distribution (see, e.g., Fishman \& Meegan 1995) would, apart
from magnetic deflection of the protons, predict a
highly isotropic UHE CR distribution as well. Since the rate of
cosmological GRBs within the field of view of the cosmic ray
experiments which detected events above $100\,$EeV is about 1
per 50 yr, a dispersion in UHE CR arrival time of at least
$50\,$yr is necessary to reconcile observed UHE CR and GRB rates.
Such a dispersion could be caused by the time
delay of protons due to magnetic deflection (Waxman 1995a,
Waxman \& Coppi 1996).

The third class of models are the so-called ``top-down'' (TD)
models. There, 
particles are created at UHEs in the first place by the decay of
some supermassive elementary ``X'' particle associated with
possible new fundamental particle physics near the grand
unification scale (Bhattacharjee, Hill, \& Schramm
1992). Such theories predict phase transitions in the early
universe which are expected to create topological
defects such as cosmic strings, domain walls and magnetic
monopoles. Although such defects are topologically stable and
would be present up to today, they could release X particles due
to physical processes such as collapse or annihilation. Among
the decay products of the X particle are jets of hadrons. Most
of the hadrons in a jet (of the order of $10^4-10^5$) are in the
form of pions
which subsequently decay into $\gamma$-rays, electrons, and
neutrinos. Only a few percent of the hadrons are expected to be
nucleons (Hill 1983). Typical features of these scenarios are thus the
predominant release of $\gamma$-rays and neutrinos and spectra
which are considerably harder than in case of shock
acceleration. For more details about these models, see,
e.g. Bhattacharjee \& Sigl 1995.

Most recently, a possible correlation of a subset of events
above $40\,$EeV among each other and with the supergalactic
plane was reported by the AGASA experiment
(Hayashida et al. 1996). Among 20 events with energy above
$50\,$EeV, two pairs with an angular
separation of the paired events of less than $2.5^\circ$ were
observed within $10^\circ$ of the supergalactic plane.
The probability for that to happen by chance for an
isotropic, unclustered distribution was given to be
$\simeq6\times10^{-4}$. A third pair was observed among 36
showers above $40\,$EeV, with a chance probability of about
$1\%$ (Hayashida et al. 1996). This suggests that the
events within one pair have been emitted by a single discrete
source possibly associated 
with the large scale structure and within a time scale of order
$2\,$yr or shorter. The fact that the lower energy events in the
pairs observed by AGASA always (albeit with very poor statistics
since only 3 pairs were observed)
arrived later suggests that the pair was produced in a burst and
the time delay is dominated by
magnetic deflection of the (charged) lower energy
particle. Furthermore, the distance to the source
cannot be much larger than $\simeq30\,$Mpc if the higher
energy event has been caused by either a nucleon, a nucleus, or a
$\gamma$-ray, since its energy was observed to be $\ga75\,$EeV
in all three pairs.

In this Letter we investigate possible consequences of a
confirmation of the above mentioned scenario of bursting sources
suggested by the
AGASA results. In \S 2 we discuss consequences for the strength
and structure of the galactic and extragalactic magnetic fields.
In \S 3, implications for the different
classes of UHE CR models currently discussed in the
literature are addressed. We summarize our findings in \S
4. Throughout the paper we use natural units with $c=\hbar=1$.

\section{Magnetic Field Constraints from Time Delay and Deflection}
Let us assume that the extragalactic magnetic field (EGMF) can
be characterized by a typical field strength $B$ and a coherence
length scale $l_c$. Over distances $r<l_c$, a relativistic
particle of energy $E$ and charge $e$ will then be deflected by
an angle $\alpha=r/r_l$ where $r_l=E/eB$ is the Larmor
radius. Assuming a random walk over distances $r>l_c$
leads to an average deflection angle $\alpha_{\rm
rms}=(r/l_c)^{1/2}l_c/r_l$. The time delay $\tau$ caused by these
deflections can then be written as
\begin{equation}
  \tau\simeq{\alpha_{\rm rms}^2r\over2}
  =9\times10^3\,\left({E\over100\,{\rm EeV}}\right)^{-2}
  \left({r\over10\,{\rm Mpc}}\right)^2
  \left({B\over10^{-9}\,{\rm G}}\right)^2
  \left({l_c\over1\,{\rm Mpc}}\right)\,{\rm yr}
  \,.\label{delay}
\end{equation}
Assuming that the correlated events have been produced in a
burst on a time scale much shorter than $1\,$yr such that the
time delay $\simeq2\,$yr observed by AGASA is dominated by the
deflection time Eq.~(\ref{delay}) for the lower energy event
with energy $E\simeq50\,$EeV,
the first equality in Eq.~(\ref{delay}) implies
\begin{equation}
  \alpha_{\rm rms}\simeq0.012^\circ\,\left({r\over30\,{\rm
  Mpc}}\right)^{-1/2}\,.\label{defextr}
\end{equation}
Thus, if the observed angular deviation of the correlated events
of about $2^\circ$ is caused by deflection and not by the finite
angular resolution (which is $\simeq1.6^\circ$), the contribution of
the EGMF to the deflection must be negligible. From this we
obtain the following constraint on the EGMF:
\begin{equation}
  B\la2.5\times10^{-12}\,
  \left({l_c\over1\,{\rm Mpc}}\right)^{-1/2}
  \left({r\over30\,{\rm Mpc}}\right)^{-1}\,{\rm G}
  \label{EGMF}
\end{equation}
Note that for a continuously emitting source the time delay
could be source intrinsic in which case we can only impose the
constraint $\alpha_{\rm rms}\la2.5^\circ$, leading to the less
stringent constraint
\begin{equation}
  B\la5\times10^{-10}\,
  \left({l_c\over1\,{\rm Mpc}}\right)^{-1/2}
  \left({r\over30\,{\rm Mpc}}\right)^{-1/2}\,{\rm G}
  \,.\label{EGMF2}
\end{equation}

The constraints Eq.~(\ref{EGMF}) and in particular
Eq.~(\ref{EGMF2}) for bursting sources are
considerably more stringent than
the existing upper limit on a coherent, all-pervading field of
$10^{-9}\,$G, coming from Faraday-rotation measurements (Kronberg
1994). If the charged particle would be produced as a heavy
nucleus, its 
charge would decrease on the way to the observer by partial or
complete photodisintegration in the CMB (Puget, Stecker, \&
Bredekamp 1976). In that case the bounds Eqs.~(\ref{EGMF}) and
(\ref{EGMF2})
would become even stronger by at least a factor $Z_f$, where
$Z_f$ is the charge of the nucleus upon arrival at the
observer.

If observed galactic magnetic fields cannot be explained
by a galactic dynamo (Kulsrud \& Anderson 1992), one might
expect protogalactic fields of strength $10^{-12}-10^{-9}\,$G
with a coherence scale of order $1\,$Mpc, depending on the way
this field is compressed during galaxy formation (Kulsrud et
al. 1995). A bound such as Eq.~(\ref{EGMF}) would then
considerably constrain such a scenario. An  all-pervading field would
have to be $\la10^{-12}\,$G, whereas stronger fields could not
permeate intergalactic space uniformly.
Correlations between UHE CR events might therefore offer
a means to constrain the EGMF in a way which is complementary to other
recently suggested methods (Plaga 1995; Lee, Olinto, \& Sigl
1995; Sigl, Lee, \& Coppi 1996a).

In case of bursting sources, we can thus assume that the
observed deflection is dominated by
the galactic magnetic field. If we assume this field to be
coherent over a scale $l_g$, its strength being $B_g$,
we obtain for $E=50\,$EeV
\begin{equation}
  \alpha\simeq1.1^\circ\,Z\left({l_g\over1\,{\rm kpc}}\right)
  \left({B_g\over10^{-6}\,{\rm G}}\right)\sin\theta\,,\label{gal1}
\end{equation}
for a nucleus of charge $Ze$, where $\theta$ is the
angle between the field polarization and the arrival direction
of the particle. In addition, since the time delay in the
galactic magnetic fields is bounded by the observed time delay
$\tau\simeq2\,$yr, we obtain from $\tau\simeq\alpha^2l_g/2$
\begin{equation}
  l_g\la1\,\left({\alpha\over2^\circ}\right)^{-2}\,{\rm
  kpc}\label{gal2}\\
\end{equation}
The numbers in Eqs.~(\ref{gal1}) and (\ref{gal2}) are quite
consistent with observational knowledge on the
galactic magnetic field parameters, $l_g\simeq$
hundreds of pc, $B_g\simeq3\times10^{-6}\,$G (Vallee 1991). In
addition, the observed
polarization of the coherent component of the galactic
field predicts the arrival directions of lower energy protons to
be of lower galactic latitude than the ones of the higher energy
particle (Sigl, Schramm, \& Bhattacharjee 1994). Within the
experimental angular resolution this is
consistent with the pairs observed by AGASA. Furthermore, using
standard values for the galactic magnetic field parameters,
Eq.~(\ref{gal1}) shows that it is unlikely that the clustered
events have been caused by heavy nuclei with $Z$ greater than a
few. The relative deflection of such nuclei would typically be
substantially larger than $1^\circ$ at the energies under
consideration.

\section{Implications for Ultra-High-Energy Cosmic Ray
Production Scenarios}

It was mentioned that two of the three pairs observed by AGASA lie
within $\simeq10^\circ$ of the supergalactic plane. That seems
to suggest an origin in some conventional sources associated
with the large scale structure such as powerful galaxy clusters
or AGN. Since no such object was identified as an obvious
source, the situation with regard to conventional shock
acceleration models remains inconclusive at the present time. It
has also been noted recently (Waxman, Fisher, \& Piran 1996)
that a strong concentration of UHE CRs towards the supergalactic
plane would be inconsistent with a correlation with the known
large scale structure.

This raises the question about the perspectives of alternative models
to explain possible correlations between events with energy
slightly below and above $60\,$EeV. In this energy range the
most readily detectable particles are $\gamma$-rays and
nucleons whose range is limited to less than
$\simeq100\,$Mpc (see, e.g., Lee 1996). The AGASA experiment is
approximately
sensitive to a cone with opening angle $\simeq45^\circ$ around
the zenith. Since under our assumptions time delays in the EGMF
are comparable to or smaller than the inverse rate of observed pairs
$\simeq1\,$yr, the rate of bursts $f_b$ causing these pairs must
obey $f_b\sim1.6\times10^{-6}\,{\rm Mpc}^{-3}\,{\rm
yr}^{-1}$. The combined integral flux above $100\,$EeV from
Fly's Eye (Bird et al. 1994) and AGASA (Yoshida et al. 1995) is
$J(100\,{\rm EeV})\simeq5\times10^{-21}\,{\rm cm}^{-2}\,{\rm
sr}^{-1}\,{\rm s}^{-1}$. From this, we can obtain a rough
estimate of the necessary energy release $E_b$ per burst,
\begin{equation}
  E_b\simeq{4\pi EJ(E)\over\lambda(E)f_b}
  \simeq6.3\times10^{49}
  \left({\lambda(E)\over30\,{\rm Mpc}}\right)^{-1}
  \left({J(100\,{\rm EeV})\cdot{\rm cm}^2\,{\rm sr}\,{\rm s}
  \over5\times10^{-21}}\right)
  \left({f_b\cdot{\rm Mpc}^3\,{\rm
  yr}\over1.6\times10^{-6}}\right)^{-1}
  \,{\rm erg}\,,\label{Eb}
\end{equation}
where $\lambda(E)$ is the attenuation length of the particle
species dominating the observed flux.

In relativistic fire ball models of GRBs the time scale for
proton acceleration is limited by the dissipation radius
$r_d\la\gamma_b^2
t_\gamma\la2.9\times10^{-3}(\gamma_b/300)^2\,{\rm yr}$, where
$\gamma_b$ is the Lorentz factor of the expanding fire ball and
$t_\gamma\sim1\,{\rm s}$ is the observed duration of the (low
energy) $\gamma$-ray burst. Thus, in these models the release
time scale of UHE CRs is indeed short compared to the time delay
in the observed pairs. However, the rate of cosmological GRBs,
$f_\gamma\simeq3\times10^{-8}\,{\rm Mpc}^{-3}\,{\rm yr}^{-1}$
(Cohen \& Piran 1995), violates the above condition
$f_b\sim1.6\times10^{-6}\,{\rm Mpc}^{-3}\,{\rm yr}^{-1}$. This
condition can only be circumvented if 
charged UHE CRs are delayed by at least $50\,$yr during
propagation, most probably by deflection in a large scale EGMF
(Waxman 1995a,b).
Confirmation of typical time delays in clustered events as small as a
few years would thus most likely rule out this type of
cosmological GRB models
as an explanation for such clusters. This is in analogy to the
fact that confirmation of recently claimed positional
coincidences between highest energy cosmic rays and strong GRBs
(Milgrom \& Usov 1995) would rule out an origin of UHE CRs in
cosmological GRBs. In that case, an association of UHE CRs with
GRBs would at best be possible if GRBs were situated in the
galactic halo (Vietri 1995b), an option which might soon be
ruled out by an increasing data set on GRBs. This could well
hint to the existence of exotic sources which are neither linked
to ordinary prominent astrophysical objects nor to GRBs.

Let us now assume that the burst sources consist of topological
defects. For example, certain classes of cosmic string loops
might collapse and release all of their energy in form of UHE
CRs within about one light crossing time $t_b$ (Bhattacharjee \&
Rana 1990). If $v$ is the symmetry breaking scale associated
with the phase transition in which the string was formed,
$t_b\simeq2.6\,(E_b/6.3\times10^{49}\,{\rm erg})(v/10^{23}\,{\rm
eV})^{-2}\,{\rm s}\ll1\,{\rm yr}$ and thus the ``burst
condition'' is fulfilled.

\begin{figure}
\epsscale{0.80}
\plotone{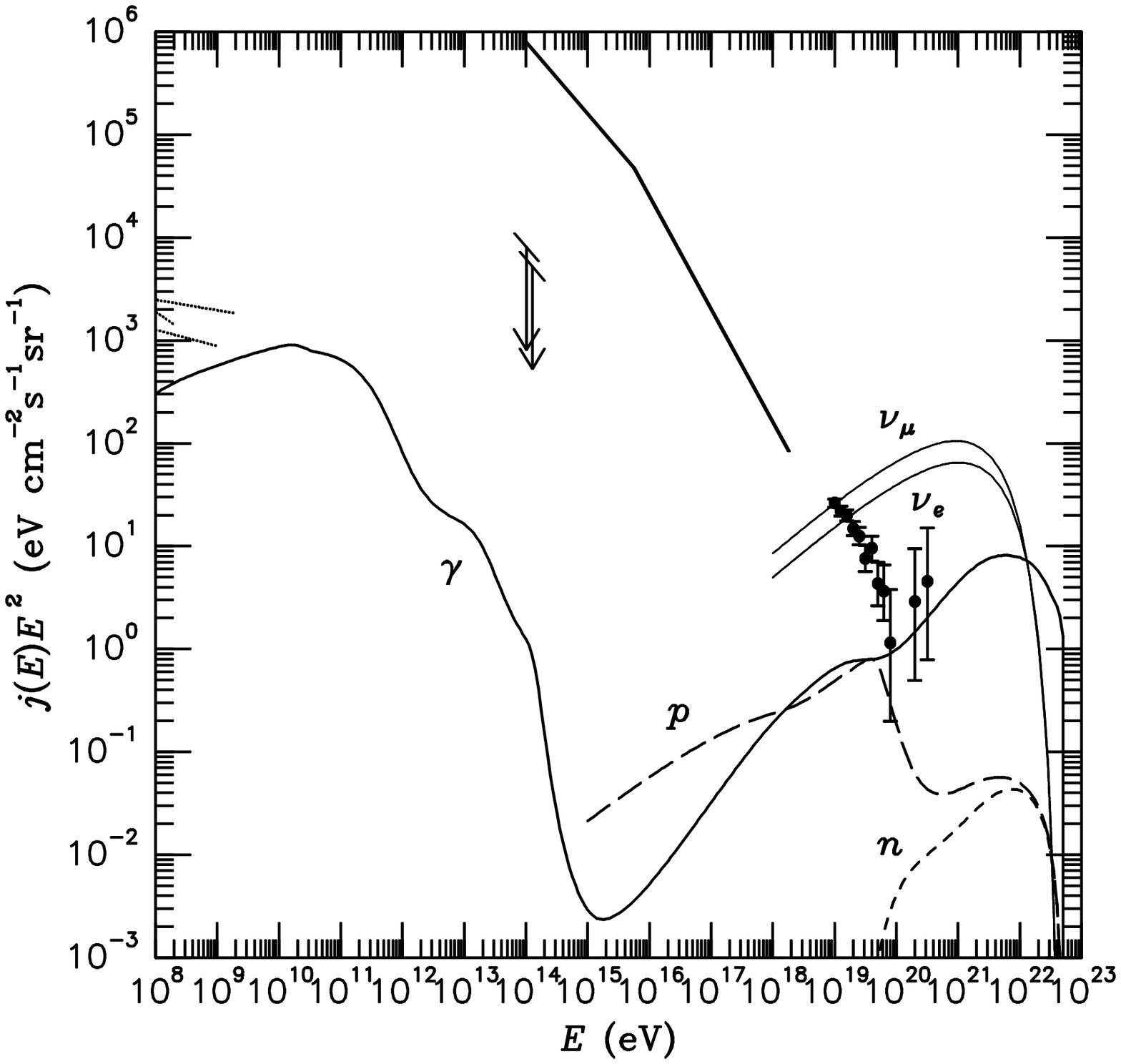}
\caption{Predictions for the differential fluxes of $\gamma$-rays (solid
line), protons (long dashed line) and neutrons (short dashed
line) above $10^{15}\,$eV and muon and electron neutrinos (thin
solid lines in decreasing order) above $1\,$EeV by a typical
TD scenario. These fluxes are time averaged or, equivalently,
for spatially uniform injection. About $3\%$ of the total energy
is injected as nucleons, $30\%$ as $\gamma$-rays, and the rest as neutrinos
with a spectrum roughly $\propto E^{-1.5}$ up to
$E=10^{23}\,$eV (for more details about the model and the
simulations see
Sigl, Lee, \& Coppi 1996a). The average modulus of the EGMF
amplitude was assumed to be $2\times10^{-12}\,$G.
Also shown are the combined data from the Fly's
Eye (Bird et al. 1993, 1994) and the AGASA
(Yoshida et al. 1995) experiments above $10\,$EeV
(dots with error bars), piecewise power law fits to the observed
charged CR flux (thick
solid line) and experimental upper limits on the $\gamma$-ray
flux below $10\,$GeV from Digel, Hunter, \& Mukherjee 1995,
Fichtel et al. 1977, and Osborne, Wolfendale, \& Zhang 1994 
(dotted lines on left margin in decreasing
order). The arrows indicate limits on the $\gamma$-ray flux
from Karle et al. 1995. Implications for the interpretation of
event clusters originating in TD induced bursts are discussed in the text.
\label{F1}}
\end{figure}

It remains to determine the UHE CR composition predicted by TD
models. We have recently performed
extensive numerical simulations for the propagation of
extragalactic nucleons, $\gamma$-rays, and electrons with
energies between $10^8\,$eV and $10^{25}\,$eV through the
universal low energy photon background (Lee 1996; Sigl, Lee, \&
Coppi 1996a,b). All relevant interactions have been taken into
account, including synchrotron loss in the EGMF of the
electronic component of
the electromagnetic cascades which result from UHE $\gamma$-ray
injection into the universal radiation background. Here, we
assume an EGMF of $2\times10^{-12}\,$G which obeys the constraint
Eq.~(\ref{EGMF}). Time averaged predictions from a representative TD model are
shown in Fig.~\ref{F1}. Since for the burst rates suggested by
the clustering observed by AGASA at any time roughly one burst
contributes to the flux above a few tens of EeV, the UHE fluxes
at these energies are representative for a typical burst induced
by a topological defect at a distance $\simeq50-100\,$Mpc.
The flux normalization was
optimized to allow for an explanation of the highest energy
events observed and corresponds to a likelihood significance for
this fit of $\simeq0.95$ above $100\,$EeV (including all bins where no
events have been detected; for details see Sigl et
al. 1995). The flux below a few tens of EeV is presumably
produced by conventional shock acceleration. The $\gamma$-ray
flux below $\sim10^{14}\,$eV only depends on the total energy
release integrated over redshift which, for a given TD model and
cosmological history of energy release, is then determined by
the flux normalization at UHEs. It can clearly be
seen that the scenario shown
in Fig.~\ref{F1} is consistent with current data and bounds on
$\gamma$-ray and UHE CR fluxes. For more details on constraints
on TD models see Sigl, Lee, \& Coppi 1996a,b; Sigl et al. 1995.
Fig.~\ref{F1} shows that events above
$\simeq80\,$EeV are predicted to be most likely $\gamma$-rays,
whereas around $50\,$EeV an approximately equal amount of
protons is expected from the TD induced bursts. About one fifth
of the total observed flux at these energies would be due to
protons from the TD induced bursts, in rough agreement with the two
observed pairs out of 20 events above $50\,$EeV (clusters of
pure $\gamma$-rays within a timescale roughly given by the
burst itself should, of course, eventually also be seen in this
scenario for sufficiently high total exposure). Since
an electromagnetic cascade particle (i.e. a $\gamma$-ray or an
electron) is deflected and delayed equally or less strongly than
a proton at the
same energy, this scenario is clearly consistent with the
discussion of the previous section. We also note that the muon
content of the showers observed by AGASA is not in
contradiction with interpreting the higher energy event in the
pairs as a $\gamma$-ray (Hayashida et al. 1996).

\section{Conclusions}
We discussed the consequences of a possible clustering of a
subset of UHE CR events above $\simeq40\,$EeV which was recently
reported by the AGASA experiment. If the observed time delay of
low relative to high energy events of $\simeq2\,$yr is typical,
the correlated events might originate in a burst on a timescale
shorter than $\sim1\,$yr, the observed time delay being caused by
deflection in magnetic fields.
If the real angular deviation between clustered events is not
much smaller than $1^\circ$ (which currently cannot be excluded
because the angular resolution of the AGASA experiment is
comparable to the observed deviation), deflection of charged
particles by the EGMF
should be negligible and can be exclusively attributed to the
galactic magnetic field, provided the charge is smaller than a
few times the proton charge. This would substantially improve
existing limits on the EGMF. Scenarios where the magnetic fields
observed in galactic disks originate from an EGMF of strength
$10^{-12}-10^{-9}\,$G with coherence length scales
of $\simeq1\,$Mpc would be constrained considerably. This could
possibly indicate that such protogalactic fields cannot be primordial,
i.e. permeate all of intergalactic space.

The typical time delay of $\simeq2\,$yr between lower and higher
energy events within a cluster suggested by AGASA is in conflict
with models which associate such events with cosmological GRBs.
Conventional shock acceleration models require identification of
a prominent astrophysical object as a source candidate within a
few degrees of the arrival directions of the events. No obvious
identification could be made for the event clusters
observed. This might hint to the operation of a TD type
mechanism where part of the UHE events would be related to new
fundamental physics near the grand unification scale. In such a
scenario, events below and above $\simeq80\,$EeV
could be mostly nucleons and $\gamma$-rays, respectively,
if the EGMF is $\la10^{-11}\,$G and the event pairs observed by
AGASA could be produced in bursts on timescales less than
$\simeq1\,$yr. This is similar to the cosmological GRB scenario,
but with a higher burst rate per volume or, correspondingly, a
lower energy release per burst. This possibility is currently
not ruled out by any data.

Future instruments in construction or in the proposal stage such
as the Japanese Telescope Array (Teshima et al. 1992), the High
Resolution Fly's Eye (Bird et al. 1995b), and the Pierre Auger Project
(Cronin 1992) will have the potential to test whether there is
significant clustering of UHE CRs. The latter experiment, with
an angular resolution of a fraction of $1^\circ$ and an energy
resolution of $\simeq10\%$, should
detect tens of event clusters per year if the clustering observed
by AGASA is real.

\acknowledgments
We thank Jim Cronin and Angela
Olinto for invaluable comments and discussions.
We also acknowledge helpful discussions with Felix Aharonian,
Pijush Bhattacharjee, Al Mann, and Paul Sommers.
This work was supported by the DoE, NSF and NASA at the University of Chicago,
by the DoE and by NASA through grant NAG5-2788 at Fermilab, and
by the Alexander-von-Humboldt Foundation. S.L. acknowledges the
support of the POSCO Scholarship Foundation in Korea.

\newpage


\begin{references}

\reference{bhs} Bhattacharjee, P., Hill, C. T., \& Schramm,
D. N. 1992, \prl, 69, 567.

\reference{br} Bhattacharjee, P., \& Rana, N. C. 1990,
Phys. Lett. B, 246, 365.

\reference{bs} Bhattacharjee, P., \& Sigl, G. 1995, \prd, 51,
4079.

\reference{bird1} Bird, D. J., et al. 1993, \prl, 71, 3401.

\reference{bird2} Bird, D. J., et al. 1994, \apj, 424, 491.

\reference{bird3} Bird, D. J., et al. 1995a, \apj, 441, 144.

\reference{bird4} Bird, D. J., et al. 1995b, in Proc. 24th
Cosmic-Ray Conf. (Rome), OG sessions, Vol. 2, 504.

\reference{be} Blandford, R., \& Eichler, D. 1987, Phys. Rep.,
154, 1.

\reference{cp} Cohen, E., \& Piran, T. 1995, \apj, 444, L25.

\reference{cronin} Cronin, J. W. 1992, Nucl. Phys. B
(Proc. Suppl.), 28B, 213.

\reference{dhm} Digel, S. W., Hunter, S. D., \& Mukherjee,
R. 1995, \apj, 441, 270.

\reference{es} Elbert, J. W., \& Sommers, P. 1995, \apj, 441,
151.

\reference{fichtel} Fichtel, C. E., et al. 1977, \apj, 217, L9.

\reference{fm} Fishman, G. J., \& Meegan, C. A. 1995, ARA\&A,
33, 415.

\reference{grei} Greisen, K. 1966, \prl, 16, 748.

\reference{hvsv} Halzen, F., Vazques, R. A., Stanev, T., \&
Vankov, H. P. 1995, Astropart. Phys., 3, 151.

\reference{haya1} Hayashida, N., et al. 1994, \prl 73, 3491.

\reference{haya2} Hayashida, N., et al. 1996, report
ICRR-361-96-12, submitted to \prl.

\reference{Hill} Hill, C. T. 1983, Nucl. Phys. B, 224, 469.

\reference{hillas} Hillas, A. M. 1984,
Ann. Rev. Astron. Astrophys., 22, 425.

\reference{karle} Karle, A., et al. 1995, Phys. Lett. B, 347,
161.

\reference{k} Kronberg, P. P. 1994, Rep. Prog. Phys, 57, 325.

\reference{ka} Kulsrud, R. M., \& Anderson, S. W. 1992, \apj,
396, 606.

\reference{kcgs} Kulsrud, R. M., Cowley, S., Gruzinov, A., \&
Sudan, R. 1996, to appear in Phys. Rep.

\reference{lee} Lee, S. 1996, report FERMILAB-Pub-96/066-A,
astro-ph/9604098, submitted to \prd.

\reference{los} Lee, S., Olinto, V. A., \& Sigl, G. 1995, \apj,
455, L21.

\reference{Meszaros} Meszaros, P. 1995, to appear in Proc. 17th
Texas Conf. Relativistic Astrophysics (NY Acad. Sci.)

\reference{mu} Milgrom, M., \& Usov, V. 1995, \apj, 449, L37.

\reference{owz} Osborne, J. L., Wolfendale, A. W., \& Zhang,
L. 1994, J. Phys. G, 20, 1089.

\reference{plaga} Plaga, R. 1995, \nat, 374, 430.

\reference{psb} Puget, J. L., Stecker, F. W., \& Bredekamp,
J. H. 1976, \apj, 205, 638.

\reference{ssb} Sigl, G., Schramm, D. N., \& Bhattacharjee,
P. 1994, Astropart. Phys., 2, 401.

\reference{sjsb} Sigl, G., Jedamzik, K., Schramm, D. N., \&
Berezinsky, V. 1995, \prd, 41, 342.

\reference{slsb} Sigl, G., Lee, K., Schramm, D. N., \&
Bhattacharjee, P. 1995, Science, 270, 1977.

\reference{slc1} Sigl, G., Lee, S., \& Coppi, P. 1996a,
report FERMILAB-Pub-96/087-A, astro-ph/9604093, submitted to
\prl.

\reference{slc2} Sigl, G., Lee, S., \& Coppi, P. 1996b,
in preparation.

\reference{teshima} Teshima, M., et al. 1992, Nucl. Phys. B
(Proc. Suppl.), 28B, 169.

\reference{Vallee} Vallee, J. P. 1991, \apj, 366, 450.

\reference{Vietri1} Vietri, M. 1995a, \apj, 453, 883.

\reference{Vietri2} Vietri, M. 1995b, \mnras, ???, ???.

\reference{wax1} Waxman, E. 1995a, \prl, 75, 386.

\reference{wax2} Waxman, E. 1995b, \apj, 452, L1.

\reference{wax3} Waxman, E., Fisher, K. B., \& Piran, T. 1996,
astro-ph/9604005, submitted to \apj.

\reference{yosh} Yoshida, S., et al. 1995, Astropart. Phys., 3, 105.

\reference{zk} Zatsepin, G. T., Kuzmin, V. A. 1966, Pis'ma
Zh. Eksp. Teor. Fiz., 4, 114 [JETP. Lett., 4, 78].

\end{references}
\end{document}